\documentclass[10pt]{article}
\usepackage[a4paper,margin=2.54cm,includefoot]{geometry}
\usepackage{amsmath}
\usepackage{cite}
\usepackage{url}
\usepackage{amsfonts} 
\usepackage{amssymb}
\usepackage{graphicx}
\usepackage{caption}
\usepackage{tikz}
\usetikzlibrary{positioning,arrows,automata,shapes,shadows}
\usepackage[caption=false]{subfig}
\usetikzlibrary{arrows,matrix}
\usetikzlibrary{decorations.pathreplacing,calc}
\usepackage{fancyhdr}
\usepackage{algorithm}
\usepackage{algorithmic}
\usepackage{comment}
\floatname{algorithm}{Protocol}
\usepackage{paralist}
\usepackage{authblk}

\newenvironment{keywords}{
       \list{}{\advance\topsep by0.35cm\relax\small
       \leftmargin=1cm
       \labelwidth=0.35cm
       \listparindent=0.35cm
       \itemindent\listparindent
       \rightmargin\leftmargin}\item[\hskip\labelsep
                                     \bfseries Keywords:]}
     {\endlist}

\newlength{\mywidth}

\title{On the Use of Key Assignment Schemes\\ in Authentication Protocols}
\author{James Alderman}
\author{Jason Crampton}
\affil{Information Security Group\\Royal Holloway, University of London}
\date{}

\begin{document}

\maketitle

\begin{abstract}

Key Assignment Schemes (KASs) have been extensively studied in the context of cryptographically-enforced access control, where derived keys are used to \emph{decrypt} protected resources. In this paper, we explore the use of KASs in entity authentication protocols, where we use derived keys to \emph{encrypt} challenges.  This novel use of KASs permits the efficient authentication of an entity in accordance with an authentication policy by associating entities with security labels representing specific services. Cryptographic keys are associated with each security label and demonstrating knowledge of an appropriate key is used as the basis for authentication.  Thus, by controlling the distribution of such keys, restrictions may be efficiently placed upon the circumstances under which an entity may be authenticated and the services to which they may gain access.

In this work, we explore how both standardized protocols and novel constructions may be developed to authenticate entities as members of a group associated to a particular security label, whilst protecting the long-term secrets in the system.  We also see that such constructions may allow for authentication whilst preserving anonymity, and that by including a trusted third party we can achieve the authentication of individual identities and authentication based on timestamps without the need for synchronized clocks.

\end{abstract}


\section{Introduction}
  \label{sect:intro}

Key Assignment Schemes (KASs) have been studied since the work of Akl and Taylor~\cite{akl:cryp83} as a means of permitting an entity to derive many cryptographic keys by combining a small number of keys in its possession with knowledge of some publicly available information. Traditionally, such schemes are used to support cryptographically-enforced access control, particularly for information flow policies; in this setting, derived keys are used to decrypt protected resources. However, we believe that KASs can also play a role in entity authentication protocols by using the derived keys as encryption keys instead. In this paper, we investigate methods by which KASs may be integrated into existing, standardized authentication protocols in order to authenticate an entity as a member of a specified group.  Associating groups with specific services can allow for more control to be exerted over the conditions under which an entity may be authenticated, such as allowing authentication only during certain time periods (during working hours, for example) or if the entity has been assigned a specific security clearance. We shall also see that a KAS can help protect the long-term secret key in a distributed system, and allow a particular form of authentication to occur whilst preserving the anonymity of entities.

This paper focuses on two-party symmetric-key authentication protocols wherein we replace the usual long-term, shared key with one derived from a KAS construction. Authentication is achieved by constructing a fresh message using this shared secret. Keys in a KAS are associated with particular security labels which could represent security classifications, time periods or geo-spatial locations for example. Thus, by making appropriate choices of labels and KAS constructions, we can require the claimant to demonstrate knowledge of keys (derived from the shared KAS secret) which satisfy an authentication policy for the system. A claimant can achieve this if and only if they have been provided with an appropriate KAS secret and is, therefore, able to derive the necessary keys. We note that the use of a KAS instantiated with labels chosen according to an authentication policy can go some way towards authenticating an entity's identity \emph{and} establishing authorization credentials simultaneously during a single protocol run. In addition, we argue that the traditional process of determining authorization uses entity authentication primarily as a means to perform a look up on the identity and the associated permissions to determine whether the claimant should be allowed to use the service in question. To this end, in this work we adapt entity authentication protocols such that they do not necessarily verify individual identity but instead demonstrate that the claimant is associated with a specified set of permissions, and do so in a manner that requires the claimant to prove that association, thereby reducing the verifier's workload.

The remainder of this paper is organized as follows: we begin by providing a brief introduction to graph-based access control policies and key assignment schemes, before discussing how several standardized protocols may be extended using a KAS to prove membership of a group associated with a set of security labels.  We then investigate novel constructions to provide authentication using timestamps, as well as how a Trusted Third Party (TTP) may be introduced to verify individual entity identities.


\section{Background}
  \label{sect:background}

First we introduce some notation to be used in the remainder of this paper. The statement $A \rightarrow B: m$ is to be interpreted as: entity $A$ sends the message $m$ to entity $B$, whilst a message of the form $\left\{m\right\}_\kappa$ means that the plaintext $m$ has been encrypted under the key $\kappa$, and $H(m)$ denotes the output of a cryptographic hash function $H$ applied to the message $m$. Finally, we write $\kappa_{A,B}$ to denote a symmetric key shared by entities $A$ and $B$, while $\tau_A$ denotes a time-stamp and $\eta_A$ a nonce (number used only once), both created by entity $A$. We write $[i,j]$ to denote the set of consecutive integers $\{i,\dots,j\}$.


\subsection{Graph-based access control policies}
  \label{sect:graph}

A \emph{partially ordered set}, or \emph{poset}, is a set $L$ equipped with a binary relation $\leqslant$ such that for all $ x, y, z \in L$ the following conditions hold: $x \leqslant x$ (reflexivity);  if $x \leqslant y$ and $y \leqslant x$ then $x = y$ (anti-symmetry); and if $x \leqslant y$ and $y \leqslant z$, then $x \leqslant z$ (transitivity).

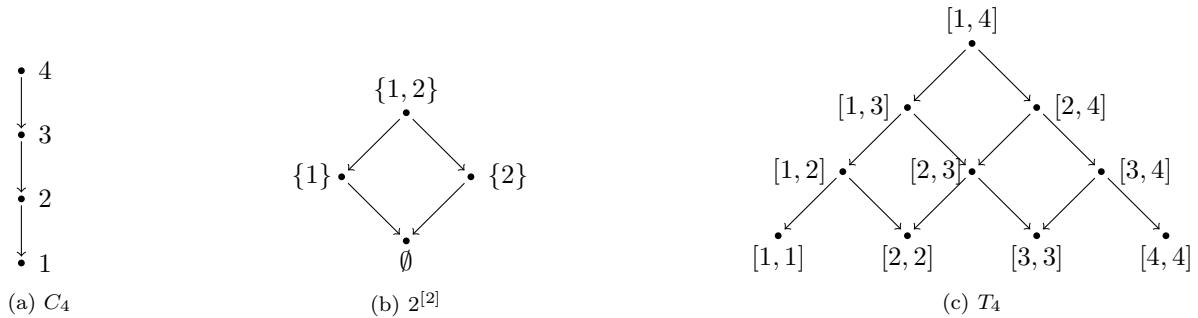
\begin{figure}
\centering
 \subfloat[][$C_4$]{
 \label{fig:hassechain}
   \begin{tikzpicture}[scale = 0.85]
   \coordinate (z4) at (0,4);
   \coordinate (z3) at (0,3);
   \coordinate (z2) at (0,2);
   \coordinate (z1) at (0,1);
   \coordinate (e4) at (0,3.9);
   \coordinate (e3) at (0,2.9);
   \coordinate (e2) at (0,1.9);
   \coordinate (e1) at (0,0.9);
   \coordinate (e8) at (0,3.1);
   \coordinate (e7) at (0,2.1);
   \coordinate (e6) at (0,1.1);
    \node[right] at (z4) {\hspace{0.1cm}$4$};
   \node[right] at (z3) {\hspace{0.1cm}$3$};
   \node[right] at (z2) {\hspace{0.1cm}$2$};
   \node[right] at (z1) {\hspace{0.1cm}$1$};
   \fill (z1) circle (1.5pt);
    \fill (z2) circle (1.5pt);
     \fill (z3) circle (1.5pt);
      \fill (z4) circle (1.5pt);
      \path[->,font=\scriptsize]
      (e4) edge node [auto] {} (e8)
      (e3) edge node [auto] {} (e7)
      (e2) edge node [auto] {} (e6);
\end{tikzpicture}
\hspace{0.001cm}
 }
 \hfill
 \subfloat[][$2^{[2]}$]{
 \label{fig:hasse4}
   \begin{tikzpicture}[scale = 0.85]
   \coordinate (z4) at (1,2);
   \coordinate (z3) at (0,1);
   \coordinate (z2) at (2,1);
   \coordinate (z1) at (1,0);
   \coordinate (e1) at (0.9,1.9);
   \coordinate (e2) at (0.1,1.1);
   \coordinate (e3) at (0.1, 0.9);
   \coordinate (e4) at (0.9, 0.1);
   \coordinate (e5) at (1.1,0.1);
   \coordinate (e6) at (1.9,0.9);
   \coordinate (e7) at (1.1,1.9);
   \coordinate (e8) at (1.9,1.1);
   \node[above] at (z4) {\vspace{0.1cm}$\left\{1,2\right\}$};
   \node[left] at (z3) {\hspace{0.1cm}$\left\{1\right\}$};
   \node[right] at (z2) {\hspace{0.1cm}$\left\{2\right\}$};
   \node[below] at (z1) {\vspace{0.1cm}$\emptyset$};
   \fill (z1) circle (1.5pt);
   \fill (z2) circle (1.5pt);
   \fill (z3) circle (1.5pt);
   \fill (z4) circle (1.5pt);
    \path[->,font=\scriptsize]
    (e1) edge node [auto] {} (e2)
    (e3) edge node [auto] {} (e4)
    (e6) edge node [auto] {} (e5)
    (e7) edge node [auto] {} (e8);
\end{tikzpicture}
 }
 \hfill
  \subfloat[][$T_4$]{
  \label{fig:hassetemp}
 \begin{tikzpicture}[scale = 0.85]
   \coordinate (z1) at (0,0);
   \coordinate (z2) at (2,0);
   \coordinate (z3) at (4,0);
   \coordinate (z4) at (1,1);
   \coordinate (z5) at (3,1);
   \coordinate (z6) at (2,2);
    \coordinate (z7) at (-1,-1);
   \coordinate (z8) at (1,-1);
   \coordinate (z9) at (3,-1);
   \coordinate (z10) at (5,-1);
   \coordinate (e1) at (1.9,1.9);
   \coordinate (e2) at (1.1,1.1);
   \coordinate (e3) at (2.1, 1.9);
   \coordinate (e4) at (2.9, 1.1);
   \coordinate (e5) at (0.9,0.9);
   \coordinate (e6) at (0.1,0.1);
   \coordinate (e7) at (1.1,0.9);
   \coordinate (e8) at (1.9,0.1);
   \coordinate (e9) at (2.9,0.9);
   \coordinate (e10) at (2.1,0.1);
   \coordinate (e11) at (3.1,0.9);
   \coordinate (e12) at (3.9,0.1);
   \coordinate (e13) at (-0.1,-0.1);
   \coordinate (e14) at (-0.9,-0.9);
   \coordinate (e15) at (0.1,-0.1);
   \coordinate (e16) at (0.9,-0.9);
   \coordinate (e17) at (1.9,-0.1);
   \coordinate (e18) at (1.1,-0.9);
   \coordinate (e19) at (2.1,-0.1);
   \coordinate (e20) at (2.9,-0.9);
   \coordinate (e21) at (3.9,-0.1);
   \coordinate (e22) at (3.1,-0.9);
   \coordinate (e23) at (4.1,-0.1);
   \coordinate (e24) at (4.9,-0.9);
   \node[left] at (z1) {\hspace{-1.1cm}$\left[1,2\right]$};
   \node[left] at (z2) {\vspace{-1.1cm}$\left[2,3\right]$};
   \node[right] at (z3) {\hspace{0.1cm}$\left[3,4\right]$};
   \node[left] at (z4) {\hspace{-1.1cm}$\left[1,3\right]$};
   \node[right] at (z5) {\hspace{0.1cm}$\left[2,4\right]$};
    \node[above] at (z6) {\vspace{0.1cm}$\left[1,4\right]$};
    \node[below] at (z7) {\vspace{0.1cm}$[1,1]$};
    \node[below] at (z8) {\vspace{0.1cm}$[2,2]$};
    \node[below] at (z9) {\vspace{0.1cm}$[3,3]$};
    \node[below] at (z10) {\vspace{0.1cm}$[4,4]$};
   \fill (z1) circle (1.5pt);
   \fill (z2) circle (1.5pt);
   \fill (z3) circle (1.5pt);
   \fill (z4) circle (1.5pt);
   \fill (z5) circle (1.5pt);
   \fill (z6) circle (1.5pt);
   \fill (z7) circle (1.5pt);
   \fill (z8) circle (1.5pt);
   \fill (z9) circle (1.5pt);
   \fill (z10) circle (1.5pt);
    \path[->,font=\scriptsize]
    (e1) edge node [auto] {} (e2)
    (e3) edge node [auto] {} (e4)
    (e5) edge node [auto] {} (e6)
    (e7) edge node [auto] {} (e8)
     (e9) edge node [auto] {} (e10)
     (e11) edge node [auto] {} (e12)
     (e13) edge node [auto] {} (e14)
     (e15) edge node [auto] {} (e16)
     (e17) edge node [auto] {} (e18)
     (e19) edge node [auto] {} (e20)
     (e21) edge node [auto] {} (e22)
     (e23) edge node [auto] {} (e24);
\end{tikzpicture}
}
 \caption{Example Hasse diagrams}
 \label{fig:hasseexamples}
\end{figure}

We may write $x < y$ if $x \leqslant y$ and $x \neq y$, and write $y \geqslant x$ if $x \leqslant y$.  We say that $x$ \emph{covers} $y$, written $y \lessdot x$, if $y < x$ and no $z$ exists in $L$ such that $y < z < x$. The \emph{Hasse Diagram} of a poset $(L,\leqslant)$ is the directed acyclic graph $(L, \lessdot)$ wherein vertices are labelled by the elements of $L$ and an edge connects vertex $v$ to $w$ if and only if $w \lessdot v$. 
We write $C_n$ to denote the chain (total order) on $n$ elements; we write $T_n$ to denote the poset $(\{[i,j] : 1 \leqslant i \leqslant j \leqslant n\},\subseteq)$ and we write $2^{[n]}$ to denote the powerset of $n$ elements. Figure~\ref{fig:hasseexamples} shows Hasse diagrams for $C_4$, $2^{[2]}$ and $T_4$.

Let $U$ be a set of entities in a distributed system, $O$ be a set of resources to which access should be restricted by a policy, and $(L,\leqslant)$ be a poset of security labels. Also, let $\lambda: U \cup O \rightarrow L$ be a labelling function assigning a security label to each entity and object. The tuple $(L,\leqslant, U, O, \lambda)$ then denotes an \emph{information flow policy} which can be represented by the Hasse Diagram of $(L,\leqslant)$.  Henceforth we shall refer to such policies as \emph{graph-based access control policies}. The policy requires that information flow from objects to entities preserves the partial ordering relation; for instance an entity $u \in U$ may read an object $o \in O$ if and only if $\lambda(u) \geqslant \lambda(o)$. Note that this statement is the \emph{simple security property} of the Bell-LaPadula security model~\cite{bell:secu73}. The enforcement of a graph-based access control policy prevents an entity assigned clearance label $x$ from accessing objects classified with label $y$ if $y > x$.  Posets of the form shown in Figure~\ref{fig:hasseexamples} have been used extensively as the basis for graph-based access control policies, notably in the Bell-LaPadula model and in temporal access control~\cite{cram:prac11}.

\subsection{Key assignment schemes}
  \label{sect:kas}

A \emph{key assignment scheme} provides a generic, cryptographic enforcement mechanism for graph-based access control policies in which a unique cryptographic key is associated to each node (representing a security label) in the graph $(L, \leqslant)$. Akl and Taylor~\cite{akl:cryp83} introduced the idea of a KAS to manage the problem of key distribution by allowing a
 trusted center to distribute a single cryptographic key, $\kappa(x)$, to each entity. The entity may then combine knowledge of this secret key with some publicly available information in order to derive additional keys $\kappa(y)$.

Henceforth, we write $\kappa_x$ to represent the cryptographic key $\kappa(x)$.
A well-known KAS construction publishes encrypted keys.  In particular, for each directed edge $(x,y)$ in the Hasse diagram, $\{\kappa_y\}_{\kappa_x}$ is published.  Then for any $x > y$, there is a (directed) path in the Hasse diagram from $x$ to $y$ and the key associated with each node on that path can be derived (in an iterative fashion) by an entity that knows $\kappa_x$.  This type of scheme has been called an \emph{iterative key encrypting} (IKE) KAS~\cite{cram:onke06}.

A fundamental security property of a KAS is that it should be secure against \emph{key recovery}~\cite{atal:dyna09}: that is, the derivation of $\kappa_y$ from a set of keys $\kappa_{x_1}, \ldots, \kappa_{x_n}$ should be possible if and only if there exists $i$ such that $\kappa_{x_i} > y$.
This property asserts that a set of users cannot recover a key for which no one of them isn't already authorized.
An IKE KAS is known to be secure against \emph{key recovery} provided the encryption function is chosen appropriately~\cite{atal:dyna09}.

Specific choices of the poset of security labels give rise to the information flow policies made famous in the Bell-LaPadula model~\cite{bell:secu73} and temporal access control policies, wherein an entity is permitted to derive cryptographic keys referring to specific intervals. For example, Figure~\ref{fig:hassetemp} illustrates a temporal poset where the leaf nodes represent specific time periods, and an entity is provided with the key relating to an interval (non-leaf node in the tree), from which it is possible to iteratively derive keys for the children. A survey of existing generic schemes is given in~\cite{cram:onke06} whilst further details of temporal access control and more general interval-based schemes can be found in~\cite{cram:prac11} and~\cite{atal:effi07}.

\section{Applying KASs to standardized authentication protocols}
  \label{sect:main}

As we have seen, KASs have been used to enforce graph-based access control policies, where a protected object is encrypted with the key associated with the object's security label and an entity may (if authorized) derive the key to decrypt the object.
However, we could also use derived keys for encrypting messages.
Given that many authentication protocols use symmetric encryption to respond to challenges, we now explore how we can use KASs to build novel authentication protocols.

\paragraph{Traditional Entity Authentication.}
Consider, for example, Protocol~\ref{prot:uni2p}~\cite[Mechanism 2]{iso:iso08}~--~a unilateral, challenge-response authentication protocol~--~in which the verifier B sends a nonce $\eta_B$ to the claimant A (message 2).\footnote{Protocols~\ref{prot:uni2p},~\ref{prot:mut3p},~\ref{prot:uni1p} and~\ref{prot:mut2p} are taken from the ISO standard~\cite{iso:iso08}. Some textual fields have been omitted from protocol descriptions in the interests of clarity and brevity.  Protocol~\ref{prot:akep} is based upon the AKEP1 protocol~\cite{bell:enti93}.}
By encrypting a plaintext that includes the nonce (message 3), the claimant demonstrates knowledge of the shared secret key $\kappa_{A,B}$ and the verifier knows that the message cannot be a replay.
A protocol in which the claimant encrypts a timestamp (Protocol~\ref{prot:uni1p}~\cite[Mechanism 1]{iso:iso08}) requires fewer messages.
However, such a protocol requires the claimant and verifier to have (loosely) synchronized clocks and for there to be some ``window of acceptability'' for timestamps.
\emph{Mutual authentication} can be achieved by requiring both parties to encrypt a message containing a nonce (Protocol~\ref{prot:mut3p}~\cite[Mechanism 4]{iso:iso08}).
A similar mutual authentication protocol using timestamps can easily be designed (Protocol~\ref{prot:mut2p}~\cite[Mechanism 3]{iso:iso08}).
Finally, in addition, a protocol may provide \emph{authenticated key exchange} (Protocol~\ref{prot:akep}) by including a session key in the verifier's response (message 2).
Notice that the protocols presented here use an authenticated encryption scheme to protect certain messages, in order to verify that only an entity in possession of a valid cryptographic key could create the message and that it has not been maliciously altered.  The protocols presented here could be modified to use a MAC (computed using the symmetric key), or other suitable cryptographic primitives if desired.

\begin{figure}[ht]
\settowidth{\mywidth}{A $\rightarrow$ B: $\left\{\eta_B, B\right\}_{\kappa_{A,B}}$\qquad}
\begin{minipage}[t]{\mywidth}
\centering
\begin{algorithm}[H]
\caption{}
\begin{algorithmic}
\STATE A $\rightarrow$ B: Hi
\STATE B $\rightarrow$ A: $\eta_B$
\STATE A $\rightarrow$ B: $\left\{\eta_B, B\right\}_{\kappa_{A,B}}$
\end{algorithmic}
\label{prot:uni2p}
\end{algorithm}
\end{minipage}
\hspace*{16pt}
\settowidth{\mywidth}{B $\rightarrow$ A: $\left\{\eta_A, \eta_B, A\right\}_{\kappa_{A,B}}$\qquad}
\begin{minipage}[t]{\mywidth}
\begin{algorithm}[H]
\caption{}
\begin{algorithmic}
\STATE A $\rightarrow$ B: $\eta_A$
\STATE B $\rightarrow$ A: $\left\{\eta_A, \eta_B, A\right\}_{\kappa_{A,B}}$
\STATE A $\rightarrow$ B: $\left\{\eta_B, \eta_A\right\}_{\kappa_{A,B}}$
\end{algorithmic}
\label{prot:mut3p}
\end{algorithm}
\end{minipage}
\hspace*{16pt}
\settowidth{\mywidth}{B $\rightarrow$ A: $\left\{\eta_A, \eta_B, A, B,   \kappa_s\right\}_{\kappa_{A,B}}$\qquad}
\begin{minipage}[t]{\mywidth}
\begin{algorithm}[H]
\caption{}
\begin{algorithmic}
\STATE A $\rightarrow$ B: $\eta_A$
\STATE B $\rightarrow$ A: $\left\{\eta_A, \eta_B, A, B,   \kappa_s\right\}_{\kappa_{A,B}}$
\STATE A $\rightarrow$ B: $\left\{\eta_B, A\right\}_{\kappa_{A,B}}$
\end{algorithmic}
\label{prot:akep}
\end{algorithm}
\end{minipage}
\hfill
\begin{minipage}[t]{\textwidth}\centering
\settowidth{\mywidth}{A $\rightarrow$ B: $\left\{\tau_A, B\right\}_{\kappa_{A,B}}$\qquad}
\begin{minipage}[t]{\mywidth}
\centering
\begin{algorithm}[H]
\caption{}
\begin{algorithmic}
\STATE A $\rightarrow$ B: $\left\{\tau_A, B\right\}_{\kappa_{A,B}}$
\end{algorithmic}
\label{prot:uni1p}
\end{algorithm}
\end{minipage}
\hspace*{16pt}
\settowidth{\mywidth}{A $\rightarrow$ B: $ \left\{\tau_A, B\right\}_{\kappa_{A,B}}$\qquad}
\begin{minipage}[t]{\mywidth}
\begin{algorithm}[H]
\caption{}
\begin{algorithmic}
\STATE A $\rightarrow$ B: $ \left\{\tau_A, B\right\}_{\kappa_{A,B}}$
\STATE B $\rightarrow$ A: $\left\{\tau_B, A\right\}_{\kappa_{A,B}}$
\end{algorithmic}
\label{prot:mut2p}
\end{algorithm}
\end{minipage}
\hfill
\end{minipage}
\hfill
\caption{Entity authentication protocols}\label{fig:authentication-protocols}
\end{figure}

\paragraph{Authentication using KASs.}
We now consider how these protocols can be modified to make use of keys derived from a KAS. We assume the existence of a \emph{graph-based authentication policy} $(L,\leqslant,U,S,\lambda)$ which we define in an analogous manner to graph-based access control policies (see Section~\ref{sect:graph}): $U$ is a set of entities, $S$ is a set of services and $L$ is a set of distinct security labels that forms a poset under the relation $\leqslant$; $\lambda: U \cup S \rightarrow L$ is a function that assigns a security label to each entity and service. Here, the term \emph{service} is used to denote the claimant's intended interaction following the successful completion of the authentication protocol: \emph{i.e.} we assume the claimant wishes to authenticate in order to be permitted to interact with the service (for example, logging in to the system or sending a document to a print server). We also assume the existence of a KAS associated with the graph-based authentication policy.

In the following protocols, we replace the symmetric key $\kappa_{A,B}$ used in the protocols in Figure~\ref{fig:authentication-protocols} with a key derived from a KAS. We assume that a trusted center initiates the setup of the system: defining a poset of security labels and a graph-based authentication policy, and instantiating the KAS construction. As an entity, $u$, joins the system, they are assigned a security label $\lambda(u)$ and given the associated cryptographic key $\kappa_{\lambda(u)}$. Henceforth, the entity may combine knowledge of this key with the public information from the KAS to derive all keys $\kappa_x$ such that $x \leqslant \lambda(u)$ -- that is, all keys that $u$ is permitted to learn in accordance with the authentication policy. Thus, the entities are assigned to groups, each associated with a particular security label and therefore permitted to interact with a specific service.

Note that we may define an equivalence relation on $U$, where $u \sim u'$ if and only if $\lambda(u) = \lambda(u')$.  The equivalence relation induces a set of equivalence classes on $U$: for each $x \in L$, the equivalence class $U_x$ denotes the set of users having security label $x$.  The aim of the claimant, then, is to demonstrate membership of the group (equivalence class) associated with a particular security label. The claimant does this by deriving and using the appropriate key to encrypt a challenge chosen by the verifier.

\subsection{Challenge-response protocols}
\label{sect:challenge}
We begin by extending Protocols~\ref{prot:uni2p} and~\ref{prot:mut3p} to make use of a KAS.
Note that correctly encrypting the challenge demonstrates knowledge of $\kappa_v$, which means that any entity with security label $w \geqslant v$ could compute this challenge.
This authentication protocol is weaker in some sense than conventional authentication protocols in that it only proves that the claimant belongs to a group $U_w$, for some $w \geqslant v$.
However, this form of authentication will suffice for many applications, in particular those applications for which no subsequent auditing or attribution of actions to individuals is required.
Note also that conventional authentication may be thought of as a degenerate case of KAS-based authentication, in which the graph is an unordered set of labels, one label per entity.

\begin{figure}[ht]
\begin{minipage}[b]{0.25\textwidth}
\centering
\begin{algorithm}[H]
\caption{}
\begin{algorithmic}
\STATE A $\rightarrow$ B: $v$
\STATE B $\rightarrow$ A: $\eta_B$
\STATE A $\rightarrow$ B: $\left\{\eta_B, B\right\}_{\kappa_{v}}$
\end{algorithmic}
\label{prot:uni2pkas}
\end{algorithm}
\end{minipage}
\hfill
\begin{minipage}[b]{0.25\textwidth}
\centering
\begin{algorithm}[H]
\caption{}
\begin{algorithmic}
\STATE A $\rightarrow$ B: Hi
\STATE B $\rightarrow$ A: $v, \eta_B$
\STATE A $\rightarrow$ B: $\left\{\eta_B, B\right\}_{\kappa_{v}}$
\end{algorithmic}
\label{prot:vchoose}
\end{algorithm}
\end{minipage}
\hfill
\centering
\begin{minipage}[b]{0.3\textwidth}
\centering
\begin{algorithm}[H]
\caption{}
\begin{algorithmic}
\STATE A $\rightarrow$ B: $v, \eta_A$
\STATE B $\rightarrow$ A: $\left\{\eta_A, \eta_B, A\right\}_{\kappa_v}, w$
\STATE A $\rightarrow$ B: $\left\{\eta_A, \eta_B\right\}_{\kappa_w}$
\end{algorithmic}
\label{prot:mut2pkas}
\end{algorithm}
\end{minipage}
\caption{Challenge-response authentication protocols using a KAS}
\end{figure}

Protocol~\ref{prot:uni2pkas} illustrates one method for incorporating a KAS into a unilateral authentication protocol. We notice that the overall structure of the protocol is very similar to the traditional case in Protocol~\ref{prot:uni2p}, however the claimant now presents the verifier with a security label, $v$, for which she wishes to be authenticated -- for
example, this label could represent credentials that $A$ claims to have, or could contain a description of the desired service (the name or type of system she is attempting to log into, for example). Instead of using a symmetric key shared by the claimant and the verifier, the claimant now derives and uses the key, $\kappa_v$, associated with the chosen security label using the KAS. Given that the claimant is provided with the cryptographic key $\kappa_{\lambda(A)}$ by the trusted center, it will be possible to prove knowledge of $\kappa_v$ if and only if they can derive $\kappa_v$ from $\kappa_{\lambda(A)}$.

The use of a nonce to demonstrate freshness is still important (even if the security label is chosen to be the current time) since all keys $\kappa_i$ for $i \leqslant \lambda(A)$ may be derived ahead-of-time once the trusted center has distributed the key
$\kappa_{\lambda(A)}$ and made the public information available. We will see in Sections~\ref{sect:time} and~\ref{sect:tsekas} some other techniques to ensure the timeliness of the response. On receipt of the initial message from the claimant with the requested security label,
the verifier must ensure that the label satisfies the authentication policy for the requested service and that the verifier himself is authorized to proceed with the authentication
(\emph{i.e.} he has security clearance at least that of the chosen label: $\lambda(B) \geqslant v$). We will shortly see a method to handle the case in which the chosen label is not suitable for the protocol to proceed, by negotiating security labels.

Protocol~\ref{prot:uni2pkas} is initiated by the claimant providing a security label on which they would like to be challenged -- we say it is an \emph{claimant-selects-label} (CSL) protocol. In some situations, it may be preferable to have a \emph{verifier-selects-label} (VSL) mechanism where the verifier chooses the necessary security label before issuing the challenge, as shown in Protocol~\ref{prot:vchoose}. For example, this may be beneficial in an environment where the required security label for all protocol runs will be equal and hence it may be more efficient for a verification server to issue the challenge than to check that labels chosen by claimants are sufficient.  On the other hand, the CSL method may be more suitable in environments where peer-to-peer
interactions are common, or the choice of services available is greater and therefore unknown to the verifier at the beginning of the protocol run.
  
Mutual authentication can be achieved in a similar fashion, as shown in Protocol~\ref{prot:mut2pkas}.  Note that both parties choose a security label (in a VSL manner) to challenge the other participant. In practice, it is likely that the labels will be chosen such that $v = w$ but it is conceivable that one entity should be of a higher security level (for example an employee submitting a report to a superior).

\paragraph{Protecting keys.}
\label{sect:protectlabelkeys}

In traditional entity authentication protocols, the claimant demonstrates knowledge of a long-term shared secret key. In the KAS authentication protocols discussed here, however, the long-term secret key is the
key issued to the entity upon joining the system, and the protocols use derived keys instead. Thus, if we restrict the challenge security labels to relate only to derived keys (and not those issued to entities), the long-term secret is never used for
encryption and is therefore protected from known-plaintext attacks. In addition, it may be advantageous to ensure that only the trusted center may have possession of the key associated with the root node of $G$, whilst entities are issued with
non-root nodes. Thus, if a entity is compromised, it may only reveal the subset of keys derived from those in its possession, while preserving the security of other keys in the KAS.

We note that Protocol~\ref{prot:uni2pkas} provides an adversary with a plaintext-ciphertext pair which may expose the derived KAS key $\kappa_v$.
This is because the final message is an encryption of the verifier's identifier (which we assume to be public knowledge) and the nonce which has been previously distributed in the clear.
To avoid this, we could encrypt the nonce in the second message using the KAS derived key and require that the claimant prove knowledge of the plaintext
value of the nonce (and thereby knowledge of the decryption key) as shown in Protocol~\ref{prot:encnonce}.

\begin{figure}[ht]\centering
\begin{minipage}{.6\textwidth}
\settowidth{\mywidth}{B $\rightarrow$ A: $\left\{\eta_B\right\}_{\kappa_{v}}$\qquad}
\begin{minipage}[t]{\mywidth}
\begin{algorithm}[H]
\caption{}
\begin{algorithmic}
\STATE A $\rightarrow$ B: $v$
\STATE B $\rightarrow$ A: $\left\{\eta_B\right\}_{\kappa_{v}}$
\STATE A $\rightarrow$ B: $\left\{\eta_B, B\right\}_{\kappa_{v}}$
\end{algorithmic}
\label{prot:encnonce}
\end{algorithm}
\end{minipage}
\hfill
\begin{minipage}[t]{\mywidth}
\begin{algorithm}[H]
\caption{}
\begin{algorithmic}
\STATE A $\rightarrow$ B: $v$
\STATE B $\rightarrow$ A: $\left\{\eta_B\right\}_{\kappa_{v}}$
\STATE A $\rightarrow$ B: $\left\{\eta_B, B\right\}_{\kappa_{s}}$
\end{algorithmic}
\label{prot:encnoncesess}
\end{algorithm}
\end{minipage}
\end{minipage}
\caption{Challenge-response authentication protocol using a KAS with protected nonce}
\end{figure}

\paragraph{Authenticated key exchange.}
Protocol ~\ref{prot:akep} may be extended to use a KAS in the obvious way, in order to distribute a session key, security label or a key relating to a specific group (or interval) from which many session keys may be derived. A session key, as defined in~\cite{bell:enti93}, should have the property that compromising one session key should not reveal information about any other session keys. Clearly, this could have implications for our system since the disclosure of a KAS key, could allow for the undesired derivation of other keys. Thus, we must ensure that if  session keys are chosen to be from a KAS construction, they should be leaf nodes (thus preventing further derivations) or the derived children of the given node must be distinct from session keys used elsewhere in the system. Also it is important to note that, if the session key is protected by the key $\kappa_v$, \emph{any} member of a group associated with a label $w \geqslant v$ could learn the key. However, by definition, all members of the group associated with label $v$ are authorized for services at that level and so session keys may be required only to protect the service from non-members.

Alternatively, a session key could be derived from information shared during a protocol run. By protecting the nonce, as in Protocol~\ref{prot:encnonce}, the participants now have a shared secret value.
This could then be used to derive additional session keys using a pseudorandom function, in much the same way as the pre-master secret is used in the SSL/TLS protocol.  Protocol~\ref{prot:encnoncesess} illustrates the use of such a derived key $\kappa_s$ to encrypt the final message.  We can extend this to a mutual authentication protocol in which both entities provide inputs (nonces) that are used in the derivation of the session key, and hence its value is not determined by any single entity.

\paragraph{Security label negotiation.}
\label{sect:negotiation}

In a CSL protocol, it is conceivable that the chosen security label may be insufficient to prove authority for the desired service, or that the verifier may not have sufficient
authority to carry out the authentication (the verifier must also at least be a member of the group associated with the challenge label in order to verify the response message; that is, we require $\lambda(B) \geqslant v$).
On the other hand, if the verifier issues the challenge label (VSL), it may be that the claimant is not able to derive the required key but would instead like to authenticate at a lower level. Therefore, it is important that the protocols allow for the
negotiation of security labels to enable both parties to agree upon a mutually known key to form the challenge.

\begin{figure}[h]\centering
\settowidth{\mywidth}{A $\rightarrow$ B: $w, \left\{\eta_B, B\right\}_{\kappa_{w}}$\qquad}
\begin{minipage}[t]{\mywidth}\centering
\begin{algorithm}[H]
\caption{}
\begin{algorithmic}
\STATE A $\rightarrow$ B: $v$
\STATE B $\rightarrow$ A: $\eta_B, \lambda(B)$
\STATE A $\rightarrow$ B: $w, \left\{\eta_B, B\right\}_{\kappa_{w}}$
\end{algorithmic}
\label{prot:neg}
\end{algorithm}
\end{minipage}
\caption{Challenge-response authentication protocol with security label negotiation}
\end{figure}

Protocol~\ref{prot:neg} illustrates such negotiation for a unilateral CSL challenge-response protocol. We assume that $v \leqslant \lambda(A)$ (else $A$ knows she can't meet the challenge) and that $v
> \lambda(B)$ (\emph{i.e.} B does not have the requisite clearance to carry out the authentication for the chosen label). We also assume that $w$ is chosen such that $w \leqslant \lambda(B)$ and $w \leqslant \lambda(A)$, and that $w$ is at least equal to
the required security clearance for the service according to the authentication policy. When presented with the challenge label $v$, the verifier responds with the chosen nonce as well as the maximal label that he is authorized to authenticate. The claimant then chooses a new label
$w$ which is acceptable to both parties -- that is, $w$ should be the greatest common descendent of $v$ and $\lambda(B)$ (computed, for example, using a least common ancestor algorithm on the `inverted' Hasse diagram of the poset of security labels \emph{i.e.}
the Hasse diagram where the $\leqslant$ edge relations have been reversed to give $\geqslant$ relations). Note that it is important that the claimant be treated as an entity of security clearance $w$ for the duration of the service, and \emph{not} of the originally requested label $v$.\footnote{ To see this, suppose otherwise that we have the labels $x \leqslant y
\leqslant z$ and that $\lambda(A) = y$ and $\lambda(B) = x$, then $A$ requests authentication on label $z$ which is refused by $B$ (since he is not of sufficient clearance) and the protocol negotiates a challenge for label $x$ which $A$ can satisfy. $A$
would then be treated as clearance $z$ (as requested) despite only proving knowledge of $\kappa_x$, which would constitute a breach of security.}  If no such descendent can be found then the protocol should be terminated since the verifier is unable to authenticate the response.

\subsection{Time-variant parameter protocols}
\label{sect:time}

We now turn our attention to authentication protocols that make use of time-variant parameters to ensure that the claimant is alive and actively participating in the protocol run.
There are two forms of time-variant parameters -- timestamps and sequence numbers.
Both require the protocol participants to share some state: in the case of timestamps, the two parties must have loosely synchronized clocks; in the case of sequence numbers, the parties must keep track of the last sequence number used in a protocol run.
In this section, we will consider the case when timestamps are used.
Our protocols can be modified very easily to accommodate the use of sequence numbers.

Protocol~\ref{prot:units} provides one-pass unilateral CSL authentication using a KAS in conjunction with a time-stamp. On receipt of the message from the claimant, the verifier must check that the ciphertext is valid and the timestamp is within an acceptable window of a synchronized clock.  Thus, the verifier is assured that the claimant has constructed a fresh message using the secret key associated to the group $v$, and hence the claimant is authenticated as a member of the group. Similarly, Protocol~\ref{prot:mutts} provides mutual authentication of both parties by symmetrically running two instances of the unilateral protocol.  As before, some negotiation of security labels may be required for authentication to terminate successfully.

\begin{figure}[ht]
\centering
\begin{minipage}[t]{0.8\textwidth}
\settowidth{\mywidth}{A $\rightarrow$ B: $v, \left\{\tau_A, B\right\}_{\kappa_{v}}$\qquad}
\begin{minipage}[t]{\mywidth}
\begin{algorithm}[H]
\caption{}
\begin{algorithmic}
\STATE A $\rightarrow$ B: $v, \left\{\tau_A, B\right\}_{\kappa_{v}}$
\end{algorithmic}
\label{prot:units}
\end{algorithm}
\end{minipage}
\hfill
\settowidth{\mywidth}{B $\rightarrow$ A: $w, \left\{\tau_B, A\right\}_{\kappa_w}$\qquad}
\begin{minipage}[t]{\mywidth}
\begin{algorithm}[H]
\caption{}
\begin{algorithmic}
\STATE A $\rightarrow$ B: $v, \left\{\tau_A, B\right\}_{\kappa_v}$
\STATE B $\rightarrow$ A: $w, \left\{\tau_B, A\right\}_{\kappa_w}$
\end{algorithmic}
\label{prot:mutts}
\end{algorithm}
\end{minipage}
\caption{Timestamp-based authentication protocols using a KAS}
\end{minipage}
\end{figure}

\section{Extended functionality using KASs}
\label{sect:extend}

In the previous section, we have seen how a KAS may be applied to standardized authentication protocols to authenticate an entity as a member of a group associated to a particular security label. We now investigate how particular choices of $(L,\leqslant)$ and minor modifications to the protocols give rise to interesting forms of authentication. We see how time-stamp based authentication can be achieved without shared state or synchronized clocks, and how a third-party may be introduced in order to authenticate individual entity identifiers and provide authenticated key exchange.

\subsection{Time-stamp based protocols without shared state} In Section~\ref{sect:time}, we saw how conventional authentication protocols that use timestamps could be modified to make use of KASs.  However, we can use a poset to model logical time.  In particular, we can use a poset of the form $T_n$ (see Figure~\ref{fig:hassetemp}) in order to verify that a claimant is permitted to authenticate at a certain point in time using one of the challenge-response protocols discussed in Section~\ref{sect:challenge}.  The verifier using a VSL protocol, for example, issues a challenge for the label associated with the current time period $t$, which the claimant may only encrypt correctly if she can derive $\kappa_t$.  She can only do this if she was issued with a key for an interval that contained $t$ by the trusted center.  This could be useful for time-bounded `guest access' to a system. Notice that the use of temporal security labels acting as logical timestamps in this fashion removes the requirement for shared state and synchronized clocks.

A natural extension is the construction of a KAS over the poset $L \times T_n$, where $L$ is the poset of security labels in the graph-based authentication policy and $T_n$ is a poset of temporal intervals.  In this construction, the number of nodes increases to $\lvert L \rvert \times \lvert T_n \rvert$ and hence storage costs increase accordingly, but the worst case derivation time is only $\lvert L \rvert + \lvert T_n \rvert$.  The advantage of using such a poset is that the claimant can present a security label $v$ and either use the current time period, $t$, or one chosen by the verifier (or equivalently a time interval), and the challenge is to derive the appropriate key $\kappa_{(v, t)} \in L \times T_n$.  Thus we can prove membership of a group \emph{and} authorization for the given time period simultaneously (for example, to enforce that a database system may be accessed only by entities with a certain clearance \emph{and} only during office hours).  

\subsection{One-round protocols without synchronized clocks}
\label{sect:tsekas}

We can construct a one-round authentication protocol (Protocol~\ref{prot:units}) but the protocol assumes that the claimant and verifier maintain (loosely) synchronized clocks.  In this section, we investigate a method by which one-round authentication may be achieved without this requirement through the use of pre-published tokens and the time-specific release of information. The advantage of such a protocol, in addition to reducing the pairwise communication costs, is that the verifier is not required to generate, store and check a unique challenge per protocol run, but rather can associate a unique challenge with each service and time period.  Also, the construction presented here leads to an efficient, centralised authentication system in which many  verifiers may publish to the same repository of tokens (for many different services, time periods and security levels) and there is a single trusted party responsible for publishing time-specific information.

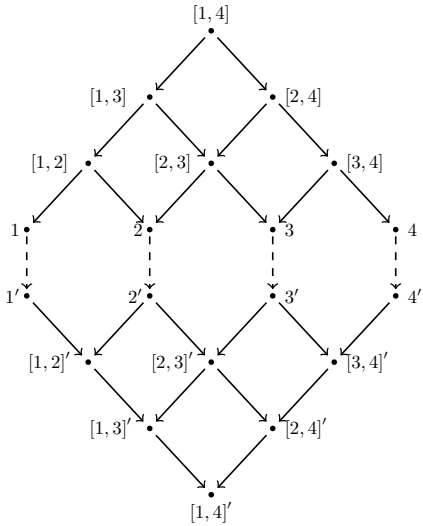
\begin{figure}[h]
\centering
\begin{minipage}[b]{.9\textwidth}
\begin{minipage}[b]{0.55\textwidth}
\begin{tikzpicture}[->,.=stealth',node distance=1cm and 9mm,semithick,auto, scale=0.7, transform shape]

  \node         (n1)                                    {};
  \node (n2)    [below left=of n1]              {};
  \node (n3)    [below right=of n1]             {};
  \node (n4)    [below left=of n2]              {};
  \node (n5)    [below right=of n2]             {};
  \node (n6)    [below right=of n3]             {};
  \node (n7)    [below left=of n4]              {};
  \node (n8)    [below right=of n4]             {};
  \node (n9)    [below left=of n6]              {};
  \node (n10)   [below right=of n6]             {};
  \node (n11)   [below=of n7]           {};
  \node (n12)   [below=of n8]           {};
  \node (n13)   [below=of n9]           {};
  \node (n14)   [below=of n10]          {};
  \node (n15)   [below right=of n11]    {};
  \node (n16)   [below left=of n13]             {};
  \node (n17)   [below left=of n14]             {};
  \node (n18)   [below right=of n15]    {};
  \node (n19)   [below left=of n17]             {};
  \node (n20)   [below left=of n19]             {};
   \node[left] at (n4) {\hspace{-1.4cm}$\left[1,2\right]$};
   \node[left] at (n5) {\hspace{-1.4cm}$\left[2,3\right]$};
   \node[right] at (n6) {\hspace{0.1cm}$\left[3,4\right]$};
   \node[left] at (n2) {\hspace{-1.5cm}$\left[1,3\right]$};
   \node[right] at (n3) {\hspace{0.1cm}$\left[2,4\right]$};
    \node[above] at (n1) {\vspace{0.1cm}$\left[1,4\right]$};
    \node[left] at (n7) {\hspace{0.1cm}$1$};
    \node[left] at (n8) {\hspace{0.1cm}$2$};
    \node[right] at (n9) {\hspace{0.1cm}$3$};
    \node[right] at (n10) {\hspace{0.1cm}$4$};

   \node[left] at (n15) {\hspace{-1.4cm}$\left[1,2\right]^\prime$};
   \node[left] at (n16) {\hspace{-1.4cm}$\left[2,3\right]^\prime$};
   \node[right] at (n17) {\hspace{0.1cm}$\left[3,4\right]^\prime$};
   \node[left] at (n18) {\hspace{-1.4cm}$\left[1,3\right]^\prime$};
   \node[right] at (n19) {\hspace{0.1cm}$\left[2,4\right]^\prime$};
    \node[below] at (n20) {\vspace{0.1cm}$\left[1,4\right]^\prime$};
    \node[left] at (n11) {\hspace{0.1cm}$1^\prime$};
    \node[left] at (n12) {\hspace{0.1cm}$2^\prime$};
    \node[right] at (n13) {\hspace{0.1cm}$3^\prime$};
    \node[right] at (n14) {\hspace{0.1cm}$4^\prime$};

\fill (n1) circle (1.5pt);
\fill (n2) circle (1.5pt);
\fill (n3) circle (1.5pt);
\fill (n4) circle (1.5pt);
\fill (n5) circle (1.5pt);
\fill (n6) circle (1.5pt);
\fill (n7) circle (1.5pt);
\fill (n8) circle (1.5pt);
\fill (n9) circle (1.5pt);
\fill (n10) circle (1.5pt);
\fill (n11) circle (1.5pt);
\fill (n12) circle (1.5pt);
\fill (n13) circle (1.5pt);
\fill (n14) circle (1.5pt);
\fill (n15) circle (1.5pt);
\fill (n16) circle (1.5pt);
\fill (n17) circle (1.5pt);
\fill (n18) circle (1.5pt);
\fill (n19) circle (1.5pt);
\fill (n20) circle (1.5pt);

  \path (n1) edge (n2)
        (n1) edge (n3)
        (n2) edge (n4)
        (n2) edge (n5)
        (n3) edge (n5)
        (n3) edge (n6)
        (n4) edge (n7)
        (n4) edge (n8)
        (n5) edge (n8)
        (n5) edge (n9)
        (n6) edge (n9)
        (n6) edge (n10)
        (n7) edge [dashed] (n11)
        (n8) edge [dashed] (n12)
        (n9) edge [dashed] (n13)
        (n10) edge [dashed] (n14)
        (n11) edge (n15)
        (n12) edge (n15)
        (n12) edge (n16)
        (n13) edge (n16)
        (n13) edge (n17)
        (n14) edge (n17)
        (n15) edge (n18)
        (n16) edge (n18)
        (n16) edge (n19)
        (n17) edge (n19)
        (n18) edge (n20)
        (n19) edge (n20);
\end{tikzpicture}
\end{minipage}
 \hfill
\settowidth{\mywidth}{TTS broadcasts: $y_{t, t^\prime}$ at time $t$\qquad}
\begin{minipage}[b]{\mywidth}
\begin{minipage}[t]{\textwidth}
\begin{algorithm}[H]
\caption{}
\begin{algorithmic}
\STATE B publishes: $\left\{\eta_B\right\}_{\kappa_{\left[t_0,t_1\right]^\prime}}$
\STATE TTS broadcasts: $y_{t, t^\prime}$ at time $t$
\STATE A $\rightarrow$ B: $\left\{B, \eta_B, \eta_A\right\}_{\kappa_{\left[t_0,t_1\right]^\prime}}$
\end{algorithmic}
\label{prot:tsekas}
\end{algorithm}
 \end{minipage}
  \hfill
  \begin{minipage}[b]{\textwidth}
\begin{algorithm}[H]
\caption{}
\begin{algorithmic}
\STATE B publishes: $\left\{\eta_B, v\right\}_{\kappa_{\left[t_0,t_1\right]^\prime}}$
\STATE TTS broadcasts: $y_{t, t^\prime}$ at time $t$
\STATE A $\rightarrow$ B: $\left\{B, \eta_B, \eta_A\right\}_{\kappa_{v}}$
\end{algorithmic}
\label{prot:tsekaslabel}
\end{algorithm}
\end{minipage}
 \end{minipage}
 \end{minipage}
  \caption{A construction for a one-round authentication protocol using pre-published tokens}
\label{fig:tse}
 \end{figure}

Figure~\ref{fig:tse} shows an example of a temporal Hasse diagram which may be used in our construction.  Notice that it may be viewed as a copy of $T_n$ reflected in the $x$-axis and joined to its reflection by a number of dashed edges.     
A KAS is instantiated over the complete Hasse diagram in which the public information used for key derivation are labels, $y_{i,j}$, associated with each edge $(i,j) \in E$. The public information associated with solid edges in the figure is published during the initialization of the system, whereas the information associated to the dashed edges will be released individually at suitable points in time.  We refer to these dashed edges as \emph{temporal edges} and to the associated information as \emph{time instant key ciphertexts} (TIKCs).  The TIKC $y_{t, t^\prime}$ is broadcast by a Trusted Time Server (TTS) at time $t$.  For example, if the KAS construction is an IKE KAS, where the public information is the set $Pub = \{\left\{\kappa_y\right\}_{\kappa_x} : y \lessdot x\}$, then the TIKC published at time $t$ will be $y_{t, t^\prime} = \left\{\kappa_{t^\prime}\right\}_{\kappa_t}$.

Protocol~\ref{prot:tsekas} illustrates how this poset may be used in a one-round authentication protocol. The verifier publishes a token, $\left\{\eta_B\right\}_{\kappa_{\left[t_0,t_1\right]^\prime}}$, in a public repository (or otherwise makes it available to the claimant).  This token is valid for a particular window of acceptance, $\left[t_0,t_1\right]^\prime$, and use of the nonce contained within it by a claimant will demonstrate liveness during that interval, as we shall see shortly.  Thus, we must ensure that only authorized entities (according to the authentication policy) are able to decrypt the token in order to retrieve the nonce. The receipt of a message containing the nonce is then sufficient to authenticate the claimant as both authorized and alive.  The verifier can make available multiple tokens associated with varying windows of acceptability depending on the security requirements for the related service.  For instance, the token for a highly time sensitive service, or where the compromise of a key will have severe consequences, may be encrypted with the key associated to a small interval in the lower half of the Hasse diagram, whilst a less critical service could use a longer window of acceptance.

\paragraph{Determining liveness.}

The token is encrypted using a key relating to a node in the lower half of the poset illustrated in Figure~\ref{fig:tse}.  Upon joining the system, the claimant is issued with a key $\kappa_v$ associated to a node $v$ in the upper half of the Hasse diagram, allowing for the derivation of keys for all descendants of $v$, that is $\left\{\kappa_w : w \leqslant v\right\}$.  Thus, an authorized entity may derive keys relating to a number of `leaf' nodes (individual time periods) of the upper poset but may not derive any keys associated with the lower half of the graph as the temporal edge labels are not yet publicly available.

Once the TTS broadcasts the TIKC $y_{t, t^\prime}$ at time $t$, a claimant that can derive the key associated with `leaf' node $t$ in the upper poset is then able to derive the key associated with the node $t^\prime$.  Notice that the nodes in the lower half of the graph can be thought of as the ``disjunction'' of individual time periods -- that is, knowledge of the key $\kappa_{t^\prime}$ for an individual time period $t^\prime$ allows for the derivation of any key associated with an interval containing $t^\prime$.  For example, to derive the key associated with the interval $\left[1,3\right]^\prime$, one must be able to derive $\kappa_{1^\prime}$ \emph{or} $\kappa_{2^\prime}$ \emph{or} $\kappa_{3^\prime}$.

By this means, a token with specified interval $\left[t_0, t_1\right]^\prime$ may not be decrypted until the TTS publishes a TIKC at a time $t \in \left[t_0, t_1\right]^\prime$ but only one such TIKC is required to successfully access the nonce.  Thus, on receipt of a valid message containing the nonce, the verifier is assured that the claimant was actively engaged in the system in order to receive the TIKC at some point during the specified window of acceptance, and this is achieved without requiring synchronized clocks.

\paragraph{Determining authorization.}

The message from the claimant is encrypted using the time-interval key $\kappa_{\left[t_0,t_1\right]^\prime}$.  Encryption is important to prevent unauthorized entities from learning the nonce and hence authenticating in a future protocol run, whilst the claimant's nonce is included to make the plaintext unique to this interaction and thus not susceptible to a replay attack. Note that the verifier must check that this nonce has not been received before in relation to this token.

Additionally, rather than using the same interval key to encrypt both the token and the claimant's message,  the token could contain a security label relating to a key that the claimant must use for encryption, as shown in Protocol~\ref{prot:tsekaslabel}.  By this means, the verifier can ask that the claimant is authorized for and alive at some point during the window of acceptance and \emph{also} authorized for a particular time period (or other security classification).  For instance, suppose the claimant is authenticating to a Ticket Generating Server which will provide a ticket with a validity lifetime equal to the window of acceptance, $\left[t_0, t_1\right]^\prime$.  Thus, Protocol~\ref{prot:tsekas} would ensure that the claimant is authorized for and alive at any point in the window of acceptance, but by including the label $\left[t_0, t_1\right]$ (relating to the upper poset) in the token (as in Protocol~\ref{prot:tsekaslabel}), the verifier also requires that the claimant has knowledge of the key $\kappa_{\left[t_0, t_1\right]}$ and is therefore authorized for the \emph{entire} interval.

Finally, it is important that the verifier maintains a list of tuples $\left\langle \eta_B, t_1\right\rangle$, where $t_1$ is the end-point of the window of acceptance, and for each protocol run must check that the end-point for the given token has not passed. This is because the TIKCs are broadcast publicly and hence malicious entities may store, or be given, TIKCs for use in future protocol runs.

\paragraph{Generalization to arbitrary posets.}

Notice that the upper half of the KAS construction in Figure~\ref{fig:tse} was chosen to be a temporal poset mirroring that of the lower half, hence requiring that the claimant is permitted to authenticate during the window of acceptance (else they could not derive the individual time period keys to which the current TIKCs relate).  However, any graph-based authentication policy is equally suitable, as long as the appropriate relations (`temporal edges') are created between the upper and lower constructions in the Hasse diagram.

More formally, suppose we have a graph-based authentication policy $(L,\leqslant)$ and the temporal poset of acceptance windows $(T_n,\supseteq)$     (the bottom half of the poset in Figure~\ref{fig:tse}), with Hasse diagrams $H(L,E)$ and $H(T_n,\supseteq)$ respectively.  We also identify a set of \emph{temporal edges}: $E^\star \subseteq L \times \left\lceil T_n\right\rceil$ where $\left\lceil T_n\right\rceil = \{[i,i]' : 1 \leqslant i \leqslant n\}$.  

During the initialization of the system, the trusted center publishes information associated with the edges in each of the two Hasse diagrams. At time $t$, the TTS publishes TIKC$_t$, the set of information that enables, for each temporal edge $(x, t^\prime) \in E^\star$, the derivation of $\kappa_{t'}$ from $\kappa_x$.  For example, if using the IKE KAS scheme, TIKC$_t  = \left\{\left\{\kappa_{t^\prime}\right\}_{\kappa_x} : (x, t^\prime) \in E^\star\right\}$.  As each TIKC is published, the two Hasse diagrams become increasingly connected by the temporal edges and the protocols defined above proceed as before.

\subsection{Verifying identity using a trusted third party}
\label{sect:ttp}

In the preceding sections, we have seen how KASs may be used to solve the generalized problem of verifying membership of a group associated with a specific security label. In this
section, we aim to achieve the special case more commonly found in entity authentication protocols -- that of proving a claim of an individual identity. An example use of such a protocol would be the act of logging into a secure database where a KAS is
used to prove clearance but knowledge of the claimant's identity is required for recording in an audit trail. Of course, this may be achieved using the techniques presented above by treating entity identifiers as security labels and partitioning entities into groups of one. Thus, we create a large KAS with a leaf or branch pertaining to each entity, but this could be prohibitively expensive due to the large number of keys and public storage required.  Instead, we now employ a Trusted Third Party (TTP) which provides additional information to the verifier that may be used to check the claimant's identity. Protocol~\ref{prot:ttp} illustrates this technique, wherein
both the entities $A$ and $B$ share a symmetric key, $\kappa_A$ and $\kappa_B$ respectively, with the TTP. Note that the trusted center responsible for the KAS and the TTP used here may belong to different security domains -- that is, we have a group
attribute provider who is authoritative on membership of security groups, and an identity provider who is authoritative on identities. In particular, the TTP may be independent of the authentication system and
entities may register their identity once with a global identity authority that acts as the TTP with many different authentication systems.

\begin{figure}
\hspace*{.1\textwidth}
\settowidth{\mywidth}{A $\rightarrow$ B: $\left\{\eta_B, B, H(\kappa_A \parallel \eta_B)\right\}_{\kappa_v}$\qquad}
\begin{minipage}[b]{\mywidth}
\begin{algorithm}[H]
\caption{}
\begin{algorithmic}
\STATE A  $\rightarrow$ B: Hi
\STATE B $\rightarrow$ TTP: $A$
\STATE TTP $\rightarrow$ B: $\left\{\eta_B, H(\kappa_A, \eta_B)\right\}_{\kappa_B}$
\STATE B $\rightarrow$ A: $v, \eta_B$
\STATE A $\rightarrow$ B: $\left\{\eta_B, B, H(\kappa_A, \eta_B)\right\}_{\kappa_v}$
\end{algorithmic}
\label{prot:ttp}
\end{algorithm}
\end{minipage}
\hfill
\settowidth{\mywidth}{TTP $\rightarrow$ B: $\left\{\eta_B, \kappa_s\right\}_{\kappa_B}$\qquad}
\begin{minipage}[b]{\mywidth}
\begin{algorithm}[H]
\caption{}
\begin{algorithmic}
\STATE A  $\rightarrow$ B: Hi
\STATE B $\rightarrow$ TTP: $A$
\STATE TTP $\rightarrow$ B: $\left\{\eta_B, \kappa_s\right\}_{\kappa_B}$
\STATE B $\rightarrow$ A: $v, \left\{\eta_B\right\}_{\kappa_v}$
\STATE A $\rightarrow$ B: $\left\{\eta_B, B\right\}_{\kappa_s}$
\end{algorithmic}
\label{prot:ttpakep}
\end{algorithm}
\end{minipage}
\hspace*{.1\textwidth}
\caption{Authentication protocols using a KAS with a TTP}
\end{figure}

In this protocol, entity authentication proceeds much as in the unilateral case, but once the claimant has initiated the protocol, the verifier seeks further information from the TTP. The TTP responds by choosing a nonce $\eta_B$ and constructing a digest using the shared key $\kappa_A$ and $\eta_B$.  The TTP concatenates this digest with the challenge and encrypts the message using the key the TTP shares with the verifier $\kappa_B$.  For concreteness, the digest in Protocol~\ref{prot:ttp} is computed as $H(\kappa_A, \eta_B)$, where $H$ is a cryptographic hash function. The inclusion of the nonce in the hash prevents the verifier from using it in a future
protocol run to impersonate the claimant, whilst the encryption prevents an adversary with a different identifier but in the same authorization group as $A$ (and hence also with access to the challenge key $\kappa_v$) from intercepting the digest and
encrypting it in the final message to successfully impersonate $A$. The verifier  forwards the nonce to the claimant along with a challenge security label $v$. The claimant responds by recreating the hash value using
the received nonce and the symmetric key $\kappa_A$ it shares with the TTP and encrypting the result with $\kappa_v$. Assuming the two hash values match, the verifier may be assured that the received digest could only have been created if the claimant had knowledge of $\kappa_A$, assumed to be known only to $A$ and the TTP. Thus, if the claimant's hash value matches that created by the TTP then the verifier may be assured both that the claimed identity is correct, and that the claimant is authorized since the message was encrypted with $\kappa_v$.

\paragraph{Ticket-based protocols.} Many entity authentication protocols in large multi-user systems  use  a TTP  to reduce the number of cryptographic keys required. Rather than every pair of entities holding a pre-agreed secret key, each entity shares a symmetric key with
the TTP and uses this to both authenticate themselves and to agree a session key for subsequent communication between the parties. In Kerberos~\cite{neu:kerb05}, for example, a user authenticates themselves to an Authentication Server (AS) using knowledge of a long-term secret (usually the user's password) and receives a ticket which can be used to access a service, or to gain
additional tickets from a Ticket Granting Server (TGS). The ticket issued authenticates the entity for a specified lifetime, which is reminiscent of a temporal KAS in which the entity can derive valid keys for a specified period of time. Modifying the Kerberos protocol to use a KAS, therefore, could achieve the same functionality whilst removing the responsibility for checking the validity of the ticket at the current time from the TGS. Instead, the entity would only be able to use the
current time period key if the key issued to her by the AS allows for its derivation, and therefore the use of that key implicitly proves authorization. In addition, proving knowledge of an interval key in a
temporal KAS construction could lead to the issuing of a ticket valid for the given interval. Early versions of Kerberos were vulnerable to a known plaintext attack~\cite{wu:area99} which revealed the user's password (the long term secret). A KAS can mitigate this risk in similar protocols by using derived keys instead of the long term secret and thus, even if such an attack succeeds, only a single key is lost and non-descendant keys remain secure.

\paragraph{Authenticated key exchange.}

Finally, note that the digest constructed by the TTP is computed over the nonce and hence is unique to this protocol run.  Therefore it could be used as a session key for subsequent communication as illustrated in Protocol~\ref{prot:ttpakep} where $\kappa_s = H(\kappa_A, \eta_B)$. The claimant proves knowledge of the group key $\kappa_v$ by decrypting the nonce and including it in the response. By encrypting the response using the session key, the verifier gains key confirmation.

\section{Related work}
\label{sect:related}

We have already discussed and referenced many of the major developments and relevant work on Key Assignment Schemes in Section \ref{sect:kas}.  In this section, we focus on related work in authentication and time-specific encryption.

\paragraph{Anonymous and membership authentication.} There has been a significant amount of work~\cite{sche:anon99, naka:anon09,ohta:memb90, tzen:secu06,bone:anon99, fuji:anon07, sant:comm98,naka:grou02,naor:deni02} relating to the concepts of \emph{anonymous authentication} and \emph{membership authentication} wherein entities are authenticated as members of a group but the verifier does not learn the individual identities. These schemes largely use public-key cryptography to demonstrate knowledge of a shared secret. Whilst anonymity was not the prime focus of our work, we note that the protocols we present in this paper provide for some degree of anonymity; specifically, users within an equivalence class $U_x$ for $x \in L$ are indistinguishable to the verifier (and, therefore, anonymous relative to the size of the equivalence class).

Previous work~\cite{ohta:memb90, tzen:secu06} has used the Akl-Taylor key assignment scheme~\cite{akl:cryp83} as a building block for anonymous authentication schemes. However, these schemes have required additional public-key mechanisms, presumably because the security of the Akl-Taylor scheme is based on the RSA problem. Our work is the first, to our knowledge, to use purely symmetric constructions.

\paragraph{Group and ring signatures.} Some membership authentication schemes use \emph{group signatures}~\cite{chau:grou91, bone:anon99, fuji:anon07, naka:grou02} or \emph{ring signatures}~\cite{rive:how01, naor:deni02} to prove knowledge of a secret known only to a group of entities.  These techniques may be used to construct public-key analogues of our (symmetric-key) protocols.  In relation to group signatures, our proposal shares the requirement of a trusted authority for initialization of the system.  However, that authority can reveal the identity of the signer, unlike our scheme(s) and ring signatures.  Moreover, ring signatures do not require a trusted authority.  However, the ease with which ring signatures can be created and the inability to trace the source of a signature makes them unsuitable for authentication in many scenarios~\cite{fuji:trac07}.  In short, we obtain the control provided by group signatures with the anonymity guaranteed by ring signatures.

\paragraph{Time-specific encryption.}
The construction presented in Section~\ref{sect:tsekas} is similar to an application of Time-Specific Encryption (TSE) suggested by Paterson and Quaglia~\cite{pate:time10}.  In their construction, built upon identity-based encryption, the information associated to temporal edges are cryptographic keys known as \emph{Time Instant Keys} (TIKs): the key $\kappa_t$ is broadcast at time $t$ and tokens are encrypted with time interval keys $\kappa_{\left[t_0,t_1\right]}$ which may only be decrypted using a key $\kappa_{t^\prime}$ for $t^\prime \in \left[t_0,t_1\right]$.  Liveness is assured by the use of a suitable TIK whilst authority is implied by the use of a symmetric key shared with the verifier.  Hence, the verifier must maintain symmetric keys with all possible claimants, which could prove intractable in large multi-user systems.  In our construction using KASs, we achieve this functionality in the symmetric setting using a relatively small number of cryptographic keys (one key per node $v \in V$) by having proof of authority come not from the claimant's identity (implied by the shared key in TSE) but by proving membership of a group associated with a certain authorization level.\footnote{Of course, if the overhead of pairwise symmetric keys between the verifier and all possible claimants is acceptable, then the final message in Protocol~\ref{prot:tsekas} may be encrypted with such a key in order to gain a stronger assurance of the claimant's identity and hence authorization.}

Notice that in TSE, the broadcasted TIK is the cryptographic key necessary to decrypt the nonce.  Thus, unauthorized recipients of this broadcast gain quite significant information -- in particular, they gain the ability to decrypt the token to recover the nonce which may provide an advantage in successfully authenticating to the verifier (in particular, they learn the contents of the claimant's message which may lead to the compromise of the symmetric key shared between the claimant and the verifier).  In our construction however, the TIKC is simply edge information that in a regular KAS would be public. Thus, a recipient of the edge information should not gain an advantage against a secure KAS construction unless in possession of a suitable cryptographic key.

\section{Summary}
  \label{sect:summ}
  
In this paper, we have presented a novel use of Key Assignment Schemes to construct entity authentication protocols.
Such protocols can be used to protect long-term secrets and to efficiently verify that a claimant satisfies an authentication policy. Example applications of such protocols include:

\begin{itemize}\setlength{\itemsep}{1pt}
\item Enforcing user clearance (for example, when accessing a secured database).
\item Authentication within a large, or rapidly changing, population where it is infeasible to maintain  a list of currently active entities, but it is possible to issue keys valid for given time periods.
\item Ticket-based authentication, where an entity is provided with a KAS key for a time interval representing a ticket lifetime. Future interactions with services require that the entity authenticate using a derived key for the current time period.
\item Authentication in subscription systems where an entity purchases a subscription for a period of time and is provided with a KAS key for that interval. When accessing the system, the entity is challenged on the key for the current time
period (the current day, for example).\footnote{Similarly, a geo-spatial KAS construction~\cite{atal:effi07} may be used such that an entity is provided with a KAS key representing geo-spatial locations. Thus, for example, a
mobile entity can be challenged on the key relating to the specific location they are attempting to access (for example, attempting to authenticate using a smart card to a secure lock within an office building).}
\item Authentication wherein entities prove authorization but wish to retain anonymity.
\end{itemize}

Further work could investigate how the techniques presented here compare to ones used for similar purposes in the public-key setting, such as ring and group signatures, particularly in terms of the applications that can be supported and the relative efficiency of the approaches.  It may also be interesting to look at mitigating denial of service attacks (DOS) on authentication servers by employing KASs in a proof of work scheme~\cite{jue:clien99}. In such a deployment, it is envisaged that a KAS be devised in which it is `moderately hard' to derive keys and thus knowledge of a key proves that significant work has been done and that the server should dedicate resources to the authentication process. The difficulty of deriving keys may be adjusted according to demand by releasing additional public information.

In addition, whilst beyond the scope of this exploratory paper, it is important to consider security definitions for KAS-based authentication and in particular whether individual security definitions for both KASs and authentication protocols hold when combined. An example problem is the notion of \emph{key indistinguishability} for KASs~\cite{atal:dyna09}. Informally, this captures the idea that an adversary should not be able to distinguish a key belonging to a KAS from a random string. However, when using such keys for authentication it becomes easy to distinguish these cases: if authentication succeeds, the adversary knows that he holds a `real' key belonging to the KAS, and thus breaks the key indistinguishability property. Finding a suitable security model for KAS-based authentication protocols is, therefore, an interesting
and important line of future work.

\bibliographystyle{plain}
\bibliography{index}

\end{document}